# Future potential for exploiting ferroelectric materials that couple pyroelectric effects with electrochemistry

Yan Zhang[1, 2], Pham Thi Thuy Phuong[3], Eleanor Roake[2], Hamideh Khanbareh[2], Yaqiong Wang[4], Steve Dunn[4], Chris Bowen[2]

1 State Key Laboratory of Powder Metallurgy, Central South University, Changsha, Hunan, 410083, China
2 Department of Mechanical Engineering, University of Bath, Bath, BA2 7AY, UK
3 Institute of Chemical Technology, Viet Nam Academy of Science and Technology, VN
4 Chemical Engineering, School of Engineering, London South Bank University, 103 Borough Road, London SE1 0AA, UK

**Introduction**

Pyroelectrics are a class of materials that convert thermal fluctuations into electric charge as a result of a change in their polarisation with temperature. Applications have focussed on their use for infrared detection and thermal imaging [1]. However, in recent years a new avenue for ferroelectric materials has emerged in applications related to electrochemical catalysis based on the pyroelectric effect. This new approach in controlling electrochemical applications involves exploiting the pyroelectric charge generated during changes in the natural ambient temperature to drive electrochemical reactions. Potential applications that have been considered to date include, air purification [2], water disinfection [3; 4], degradation of water pollutants [5-16], and water splitting for hydrogen generation [17-26].

For conventional ferroelectric applications such as piezoelectric and pyroelectric transducers that act as pressure and thermal sensors, ultrasonic devices or actuators, significant effort has been undertaken to develop ferroelectric materials with a high Curie temperature ($T_c$)[27; 28]; often well above ambient conditions ( >>100 °C). In contrast, the use of ferroelectrics with a low $T_c$ (e.g. <100 °C) is often restricted for transducer applications due to a lack of thermal stability as a result of the phase transition from a ferroelectric non-centrosymmetric crystal structure to a paraelectric centrosymmetric state above the $T_c$, whereby the polarisation level and the ferroelectric, piezoelectric and pyroelectric properties are lost.

The new and emerging area which combines pyroelectric effects with electrochemistry provides a new avenue for the low Curie temperature materials since pyroelectric properties are often maximised near the $T_c$ due to the intrinsically large change in polarisation with temperature [29]; therefore it can be assumed that any polarisation-driven pyro-electrochemical process would be greatly improved. For example, an increase in other properties such as piezoelectric coefficients, permittivity, and electrocaloric properties near the $T_c$ is common. However, current experimental work has yet to demonstrate the benefits of operating temperature near $T_c$ for pyroelectric catalysis, although a small number of low $T_c$ (< 100 °C) pyroelectrics have been successfully explored [12; 22].

This Future Energy article therefore examines this new intriguing area of research by firstly introducing the concept of using pyroelectric materials for electrochemical applications, it will then explore the potential benefits of using low $T_c$ ferroelectrics, followed by an overview of recent progress and a discussion of future opportunities for the research area.

**Combining Pyroelectrics with Electrochemical Applications**

In the 21 non-centrosymmetric crystalline family, there are only 10 polar point groups which correspond to pyroelectric materials. A spontaneous polarisation ($P_s$) is present in such pyroelectrics, in the absence of external electric field, and the pyroelectric response originates from a change in the polarisation with a change in temperature. An example of the pyroelectric effect in response to a thermal fluctuation is shown in **Schematic 1**. If we consider a pyroelectric at a temperature below $T_c$, the material is non-centrosymmetric; for example the tetragonal structure in **Schematic 1** where there is a spontaneous polarisation. This results in compensation/screening charges being attracted to the opposing sides of the material perpendicular to $P_s$ direction. As the material is heated ($dT/dt > 0$) at a relatively low temperature ($T<T_c$), the level of $P_s$ gradually decreases due to a loss of dipole orientation as a result of increased thermal vibrations; this is indicated as Region I in **Schematic 1**. This change in polarisation leads to a release of compensation charges on the two polar surfaces of the material and a current flow under short circuit conditions, or a potential difference under open circuit conditions. Similarly, on cooling the pyroelectric material ($dT/dt < 0$ and $T<T_c$), the reduction in temperature leads to an increase in polarisation level and the compensation charges redistribute to the surface, leading to a current or electric field in the reverse direction.

In the example above, we have considered a thermal fluctuation well below the $T_c$, since above $T_c$ the crystal structure becomes centrosymmetric (a cubic structure in Schematic 1) and the polarisation level reduces to zero and the material is no longer pyroelectric; indicated by Region II in **Schematic 1**. The pyroelectric coefficient (C m$^{-2}$ K$^{-1}$) is a measure of the charge released and is the gradient of the polarisation versus temperature curve in **Schematic 1**, which is defined as:

$$p = dP_s/dT \qquad (1)$$

We can see that the pyroelectric coefficient rises with increasing temperature as a result of the increased dipole and domain motion, and reaches a maximum when T ~ $T_c$ in Region III. Above the $T_c$ (Region II) both the pyroelectric coefficient level and polarisation fall to zero since the material is no longer ferroelectric. Therefore, for conventional applications, where the material is a bulk multi-domain ferroelectric with a remnant polarisation induced by a poling process, the operating temperature for a sensor or actuator is often limited to approximately one-half of $T_c$ (in unit of °C) to avoid a loss of polarisation [30].

It is relatively simple to estimate the level of charge and potential that is generated by a pyroelectric element. For an unclamped pyroelectric material, the pyroelectric charge (*Q*) generated for a temperature change (*dT*) is:

$$Q = p\, A\, dT \tag{2}$$

where *p* is the pyroelectric coefficient of the pyroelectric material and *A* is the surface area. The potential, *dV*, developed across a pyroelectric of thickness, *h*, at open circuit is given by:

$$dV = (p\, h) / (\varepsilon_{33}\, dT) \tag{3}$$

where $\varepsilon_{33}$ is the material permittivity at constant stress. It is of interest to note that the charge is proportional to the surface area (**Eqn. 2**) while the electric potential is proportional to its thickness (**Eqn. 3**), indicating a need to tune the geometry of the pyroelectric to optimise the pyroelectric performance; for example 1.23 - 1.5V is the critical potential for water splitting.

The generated electrical charge and electric potential has led to interest in using pyroelectrics as a power source for electrochemical reactions in terms of hydrogen generation, bacteria elimination, water treatment and air purification; this is indicated in the lower images of **Schematic 1** and an overview of coupling energy harvesting techniques with the electrochemical systems has been introduced by Zhang et al [31]. **Table 1** also highlights the range pyroelectric materials that have currently been adopted for electrochemical applications, which include those based on toxic lead-based materials, materials with a relatively low polarisation, low pyroelectric activity or a high Curie temperature ($T_c$>100°C) which makes transitions above and below $T_c$ in water impossible. Therefore, in addition to high polarisation, $P_s$, large surface area $A$ and optimum thickness $h$ which are beneficial for maximising the pyroelectric output, we will see in the next section that pyroelectric materials with low $T_c$ (<100 °C) offer a more practical solution to achieve large changes in polarisation, as most of the reactions take place at near ambient conditions or are in direct contact with water [32; 33]. The electrochemical application areas in Table 1 include water splitting, air purification, water disinfection and treatment and cathodic protection, which are now described.

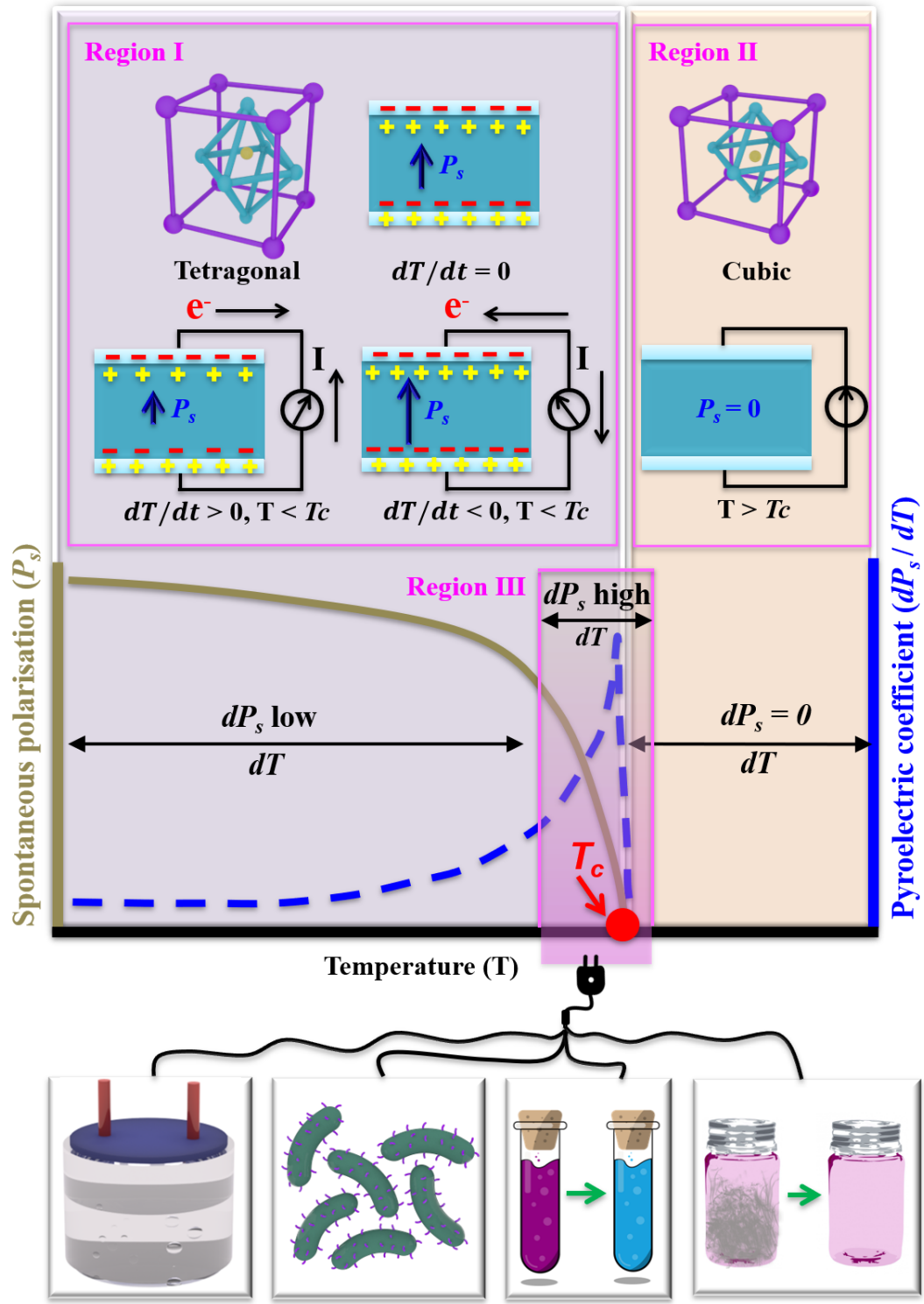

Schematic 1 The pyroelectric effect coupled with electrochemical reactions. Note the regions I (T<$T_c$, ferroelectric and tetragonal), II (T>$T_c$, paraelectric and cubic) and III (T ~ $T_c$, in the vicinity of the phase transition). Lower images indicate potential applications.

Table 1 Summary of pyroelectric materials used for electrochemical applications.

| Pyroelectric material | Nature of the work | Curie temperature,[a] $T_c$, °C | Working temperature range,[a] °C | Spontaneous polarisation,[a] $P_s$, μC/cm² | Relative permittivity at 1 kHz,[a] $\varepsilon_{33}^T$ | Pyroelectric coefficient,[a] $p$, μC/(m² K) | Pyro-figure of merit, $F_E = \frac{p^2}{\varepsilon_0 \varepsilon_{33}^\sigma}$, pJ/m³ K²,[b, c] | Application area |
|---|---|---|---|---|---|---|---|---|
| PbTiO₃ [2; 17-19] | Modelling | 495 | Below and above $T_c$ | 80 | 1200 | 100 | 0.94 | NO$_x$ reduction, CO oxidation, SO₂ oxidation, H₂ generation |
| PMNPT, PZT, BT, ZnO, PVDF, LNO, LTO [24] | Modelling | - | Only temperature gradient of 0.5 K·s⁻¹ to 1000 K·s⁻¹ | - | - | - | - | H₂ generation |
| BaTiO₃ [9; 10; 20; 25] | Experiment | 120 | 25 - 66 | 10 | 1200 | 200 | 3.77 | H₂ generation |
| | | | 30 - 47 | | | | | Rhodamine B degradation |
| | | | 30 - 52 | | | | | Rhodamine B degradation |
| | | | 30 - 54 | | | | | Rhodamine B degradation |
| PbZr$_x$Ti$_{1-x}$O₃ [5; 15; 21; 26] | Experiment | 230 - 350 | 85 - 87 | 35 | 1600 - 3800 | 268 - 390 | 2.14 – 5.07 | H₂ generation |
| | | | 28 - 60 | | | | | Rhodamine B degradation |
| | | | 22 - 37 | | | | | Rhodamine B degradation |
| LiNbO₃ and LiTaO₃ [3] | Experiment | 1140 and 607 | 20 - 45 | 0.8 and 1.7 | 28.7 and 47 | 83 and 176 | 27.12 and 74.47 | Escherichia coli disinfection |
| BiFeO₃ [34] | Experiment | 830 | 27 - 38 | 60 | 200 | 90 | 4.58 | Rhodamine B, Methyl orange, Methyl blue degradation |
| NaNbO₃ [13; 14] | Experiment | 350 | 23 - 50 | 2×10⁻⁴ | 170 | 185 | 22.75 | Rhodamine B degradation |
| | | | 15 - 50 | | | | | |
| ZnO [8] | Experiment | 147 | 22 - 62 | 5.8 | 11 | 9.4 | 0.91 | Rhodamine B degradation |
| PVDF | Experiment | 80 | 20 - 34 | 3 | 9 | 27 | 1.16 | Cathodic protection |
| Ba$_{0.7}$Sr$_{0.3}$TiO₃ [12; 35] | Experiment | 32 | 25 - 50 | 5 | 4500 | 1560 | 61.11 | H₂ generation, Rhodamine B degradation |

[a] Approximate values. [b] $\varepsilon_0$ is the permittivity of free space of 8.85 pF/m. [c] Estimate based on the experimental data reported.

**Recent Progress in Water Splitting and Water Treatment**

Research to date has involved both modelling and experimental efforts. The first evidence on the polarisation driven pyroelectric water splitting was explored by pioneering modelling in 2016 [2; 18; 19], which considered a $PbTiO_3$ ferroelectric surface which is a high $T_c$ pyroelectric with a $T_c$ ~495 °C, with further developments of modelling approaches in 2019 [24]. This work demonstrated that it is thermodynamically possible to split water into oxygen and hydrogen by thermally cycling the material surface above and below its $T_c$ when in contact with water molecules [18], a similar mechanism (**Figure 1A**) was also proposed to enable air purification by the decomposition, or oxidation, of a gaseous pollutants [2].

Experimentally, there are two potential configurations to couple pyroelectric effects with electrochemistry. This involves what we define as *external* [5; 6; 21; 26] or *internal* [20; 22; 23; 25] pyro-electrolytic systems, where the pyroelectric material is positioned either outside or inside the electrolyte, respectively.

In an *externally* positioned system, as shown in **Figure 1B** [21; 26], the pyroelectric material or element is not in direct contact with the electrolyte and is used simply as a thermal energy harvester which operates as charge source that drives an electrochemical reaction.   In this case the charge and potential difference as a result of a temperature change and can be estimated using **Equations 2** and **3**. One advantage of this approach is that since current will flow in opposite directions during the heating and cooling cycles, a rectification circuit can be used maintain a constant polarity for the separation of the negative charge and positive charge (e.g. $H^+$ and $O^{2-}$ for water splitting), leading to hydrogen production or dye decomposition at separate electrodes. While this approach is simple, the limited surface area of a bulk pyroelectric element (see **Equation 2**) leads to a restriction in the level of charge that can be generated. In addition, since the material is often a polycrystalline, bulk multi-domain material where the polarisation is achieved by a poling process to align ferroelectric domain throughout the whole material, the system much be operated in Region I of **Schematic 1** to maintain its polarisation.

A contrasting approach is an *internal* system whereby finely dispersed pyroelectric particulates are suspended in the electrolyte to provide a greatly improved pyroelectric surface area and total charge. As a result, recent work [3-16; 20; 22-25] has suspended fine scale pyroelectric particles directly into the electrolyte and subjected the system to a thermal fluctuation in a chemical reactor; for example for $H_2$ generation [20; 22-25] or water treatment studies [3-16]. An example is shown in **Figure 1C,** which shows a ferroelectric $NaNbO_3$ nanofiber subjected to heating and releases surface charge. Since the particles are effectively single domain or are simply multi-domain particles in direct contact with the electrolyte there in no need for poling to align domains of all the particles. As a result, there is potential for thermally cycling the material above and below $T_c$ and exploiting the large changes in polarisation at the Curie point (Region III) to increase the charge released from the surface during the ferroelectric to paraelectric phase transition [36].

There remain challenges for this technology, for example, in the absence of any dispersive regulation the powder may sediment to the base of the reactor, resulting in the reduced effective surface area. Furthermore, the complexity to conduct the powder collection and replacement, together with potential pH changes of the electrolyte resulting from partially dissolved powders and electrolyte contamination, especially when nano-sized powders are used in an effort in maximise the surface area.

There is also potential for multi-functional or hybrid approaches. For example, the spontaneous polarisation of a pyroelectric and internal field may also allow efficient separation of the electro-hole pairs during photocatalysis, and effort has therefore been made to couple semiconductors (ZnO [25]) and noble metals (Ag [10; 12], Pd [4; 10]) with a pyroelectric material to create heterostructures, leading to the improvement in catalytic activity for dye decomposition. An example is shown in **Figure 1D** which shows silver particles on a barium titanate ($BaTiO_3$) ferroelectric particles subjected to heating and cooling. In addition, potential hybrid power sources by integration of the pyroelectric/triboelectric technique [5] and the pyroelectric/piezoelectric effect [6; 13] are being considered build a multi-functional system in response to both thermal fluctuation and mechanical vibration for both external [5; 6] and internal [13] approaches.

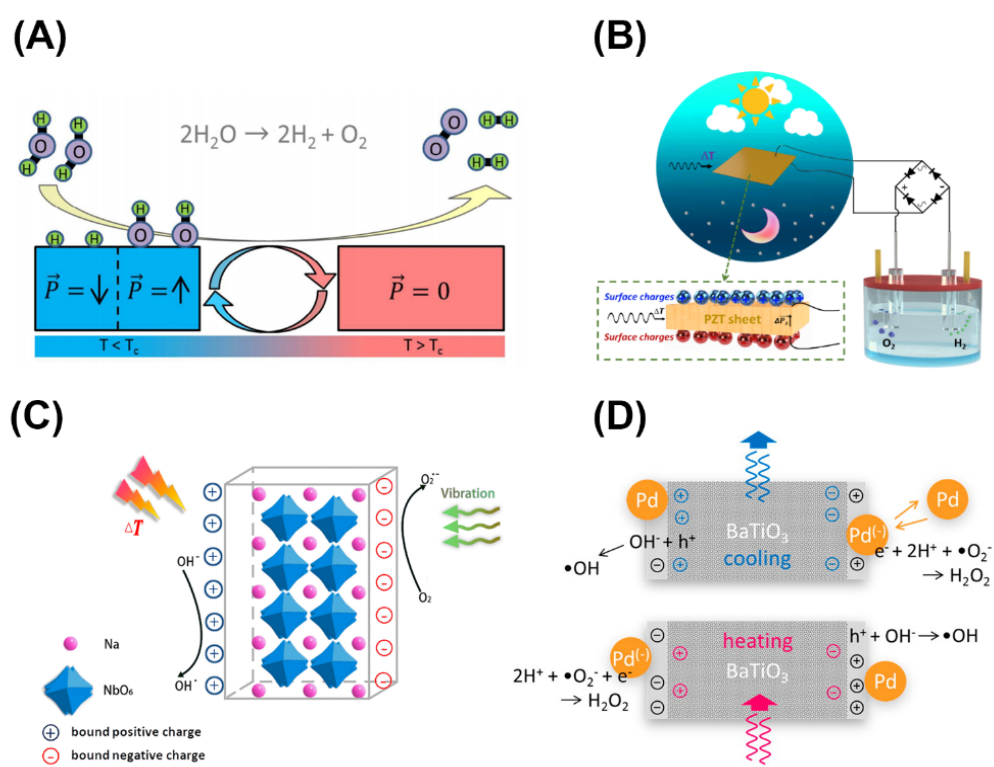

Figure 1 Mechanism and configuration of pyroelectric-electrochemical applications. (A) cyclic catalytic system for water splitting based on the pyroelectric effect [18], (B) Schematic of pyroelectric as an external source for water splitting [26], (C) Schematic of mechano-/pyro- bi-catalytic mechanism of $NaNbO_3$ nanofiber [13]. (D) Principle scheme on pyrocatalysis of $BaTiO_3$@Pd nanoparticles [4],

**Conclusions and Future Perspectives**

As an emerging approach for driving electrochemical reactions using pyroelectric materials, the presence of thermal fluctuations and/or transient waste heat in the environment has the potential to be the primary thermal input for driving the change in polarisation to release charge for such reactions. There are a number of potential avenues to explore, including:

(i) Fine scale nano-sized powders offer an opportunity to maximise the surface area and total surface charge when the pyroelectric particulates are in direct contact with the electrolyte.

(ii) There is potential to maximise performance using low $T_c$ materials (< 100 °C) as a result of the large changes in polarisation due to a transition from a low temperature ferroelectric state to high temperature paraelectric state (Region III in Schematic). To date, limited attention has been paid to the application of the low $T_c$ pyroelectrics, especially the influence of operating temperature on the pyroelectric-electrochemical performance.

(iii) Potential for creating composite mixtures with a range of Curie temperatures to harvest a range of temperature fluctuations.

(iv) There remains scope to futher increase performance, such as the creation of pyroelectric particulates combined with thermally/electrically conductive carbon or metals, combining with co-catalysts and/or exploiting plasmonic effects during heating by light.

(v) Improvement of dispersion of the particles in the electrolyte through stirring or using dispersants to increase of the effective surface area.

(vi) There is potential for froming highly porous pyroelectrics which can provide a high surface area and reduced permittivity and heat

capacity. Nanostructured ferroelectrics can also balance the high surface area and tailor the Curie temperature.

(vii) There is a need to build on existing models [17-19; 24] to understand fundamental mechanisms and provide improved estimates of efficiencies.

(viii) Future experimental work to demonstrate electrochemical applications, such as hydrogen production, air purification, bacterial disinfection, water treatment provide research space for harvesting thermal energy; including methods to achieve high frequency thermal fluctuations around phase transitions.

**Acknowlegements**

We acknowledge ERC project (ERC-2017- PoC-ERC-Proof of Concept, grant no. 789863), the Academy of Medical Sciences GCRF fund (GCRFNGR2 - 10059), and the Leverhulme Trust (RGP-2018-290). Dr Y.Z. also would like to acknowledge the support from the State Key Laboratory of Powder Metallurgy, Central South University, Changsha, China.